\documentclass[12pt]{article}
\usepackage{amssymb,graphicx,color,epsfig,psfig}
\renewcommand{\theequation}{\thesection.\arabic{equation}}

\newcommand{\be}{\begin{equation}}
\newcommand{\ee}{\end{equation}}
\newcommand{\bear}{\begin{eqnarray}}
\newcommand{\eear}{\end{eqnarray}}
\newcommand{\ba}{\begin{array}}
\newcommand{\ea}{\end{array}}
\newcommand{\nn}{\nonumber}

\begin{document}

\begin{flushright}
{\tt hep-th/0601236}
\end{flushright}

\vspace{5mm}

\begin{center}
{{{\Large \bf Late Time Behaviors of \\ an Inhomogeneous
Rolling Tachyon}}\\[14mm]
{O-Kab Kwon}\\[2.5mm]
{\it BK21 Physics Research Division and Institute of
Basic Science,\\
Sungkyunkwan University, Suwon 440-746, Korea}\\
{\tt okab@skku.edu}\\[4mm]
{Chong Oh Lee}\\
{\it Basic Science Research Institute, Chonbuk National University,
Chonju 561-756, Korea}\\
{\tt cohlee@chonbuk.ac.kr} }
\end{center}

\vspace{10mm}

\begin{abstract}
We study an inhomogeneous decay of an unstable D-brane
in the context of Dirac-Born-Infeld~(DBI)-type effective action.
We consider tachyon and electromagnetic fields
with dependence of time and one spatial coordinate,
and an exact solution is found under an exponentially decreasing
tachyon potential, $e^{-|T|/\sqrt{2}}$, which is valid for the
description of the late time behavior of an unstable D-brane.
Though the obtained solution contains both time and spatial dependence,
the corresponding momentum density vanishes over the entire spacetime
region.
The solution is governed by two parameters.
One adjusts the distribution of energy
density in the inhomogeneous direction, and the other interpolates
between the homogeneous rolling tachyon and static configuration.
As time evolves, the energy of
the unstable D-brane is converted into the electric flux and tachyon
matter.

\end{abstract}
\newpage

\section{Introduction}

Time dependent process in string theory has been intensely studied
in recent years.
Assuming that an unstable D-brane decays homogeneously, the whole
decay processes, in the vanishing string coupling limit $g_s\to 0$,
can be described by the marginally deformed boundary
conformal field theory~(BCFT)~\cite{Sen:2002nu}.
The main results of this time dependent solution, referred as
rolling tachyon, indicate that, according to time evolution,
the energy density remains
constant but the pressure goes to zero asymptotically.

On the other hand,
spatial inhomogeneity has been another important issue.
In particular, much works has been studied on
tachyon solitons, such as tachyon
kinks~\cite{Sen:2003tm,Lambert:2003zr,Brax:2003rs,Kim:2003in,Kim:2006mg}
and vortices~\cite{Sen:2003tm, Kim:2005tw}.
These solitons are interpreted as
the lower dimensional D-branes on the worldvolume of the original
unstable system.
Thus in order to see the dynamical formation of the lower dimensional D-branes,
it is indispensable to take into account the spatial inhomogeneity
in the decay process of an unstable system.

The rolling of tachyon, which is inhomogeneous along one spatial
direction, was considered in
BCFT~\cite{Sen:2002vv,Larsen:2002wc,Rey:2003zj}. The late time
behavior of the resulting energy-momentum tensor is qualitatively
different from the case of the homogeneous rolling tachyon. The
relevant components of the energy-momentum tensor exhibit
singularities at spatially periodic locations within a finite
critical time. These spatial singularities were interpreted as the
codimension-one D-branes~\cite{Sen:2002vv,Larsen:2002wc,Rey:2003zj}.
This subject was also considered in the boundary string field
theory~\cite{Ishida:2003cj} or the DBI-type effective field
theory~\cite{Felder:2002sv,Mukohyama:2002vq,Berkooz:2002je,Cline:2003vc,
Kwon:2003qn,Felder:2004xu,Barnaby:2004nk,Panigrahi:2004qr,
Saremi:2004yd,Canfora:2005mt}.
The Ref.~\cite{Felder:2002sv} showed that the inhomogeneous
solutions with a runaway tachyon potential formed caustics with
multi-valued regions beyond a finite critical time, and proposed
that in the presence of caustics, the higher derivatives of the
tachyon field blow up. For this reason the DBI-type effective action
is not reliable after the formation of caustics, since it was
proposed as an effective action for the tachyon field in string
theory where the higher derivatives of the tachyon field are
truncated~\cite{Garousi:2000tr}.

Another interesting aspect in the decay process is the dynamics at
the bottom of the tachyon potential. This process has two kinds of
decay products, which carry the effective degrees of freedom of the
original unstable D-brane, such as energy-momentum and fundamental
string charge~\cite{Yi:1999hd}. They are called tachyon
matter~\cite{Sen:2002nu,Sen:2002an} and string
fluid~\cite{Gibbons:2000hf}. In the tachyon vacuum, the dynamics of
the system is characterized as the two degrees of
freedom~\cite{Gibbons:2002tv,Kwon:2003qn,Yee:2004ec}. The main
purpose of this paper is to explore the formation of these final
states in terms of the inhomogeneous tachyon and electromagnetic
fields at late time.

Let us consider the DBI-type tachyon effective action with gauge field
interactions~\cite{Garousi:2000tr} described by
\bear \label{dbia}
S = \int d^{p+1} x {\cal L}=
- {\cal T}_p \int d^{p+1} x\, V(T) \sqrt{-X},
\eear
with
\bear
X &\equiv& \det X_{\mu\nu}
=\det (\eta_{\mu\nu}
      + \partial_\mu T \partial_\nu T +  F_{\mu\nu}),
\label{detX} \\
F_{\mu\nu}&=& \partial_\mu A_\nu - \partial_\nu A_\mu, \quad (\mu,
\nu = 0,1, \cdots, p), \nn
\eear
where $A_\mu$ is an U(1) gauge field, $V(T)$ is a runaway tachyon potential,
and ${\cal T}_p$ the tension of an unstable D$p$-brane.
This action (\ref{dbia}) is expected to provide a good description
of an unstable D-brane
in the case that the tachyon field $T$ is large,
and the higher derivatives of $T$ are small.
As we have seen in the rolling tachyon solution~\cite{Sen:2002nu},
the tachyon field $T$ goes to infinity at late time
of the D-brane decay process.
Thus DBI-type effective action describes well the late
time behaviors of the process.
However, once we take into account the inhomogeneity
of the tachyon field without gauge field interactions,
as mentioned before,
DBI-type effective action becomes inadequate in a finite
critical time in describing the dynamics of an unstable
D-brane~\cite{Felder:2002sv,Mukohyama:2002vq,Berkooz:2002je,
Cline:2003vc,Felder:2004xu,Barnaby:2004nk,Saremi:2004yd}.
Though we include the constant electromagnetic fields,
the singularity we encounter seems unavoidable~\cite{Rey:2003zj}.

In this paper, we suggest that the roles of the spacetime
dependent electromagnetic
fields are nontrivial on unstable D-brane system.
We assume that the tachyon and electromagnetic fields depend on
time and one spatial coordinate under an exponentially decreasing
tachyon potential, $e^{-|T|/\sqrt{2}}$.
We find an exact solution in a frame which gives vanishing momentum
density provided by an appropriate electromagnetic fields.
The solution represents periodic profile along the spatial direction and
it involves the  interesting and inaccessible regions alternatively.
In the interesting regions, the solution has no singularities in time direction,
and describes the late time behaviors of an unstable D-brane decay.

In section 2, we describe the calculation of the
exact solution for the tachyon and electromagnetic fields.
In section 3, we analyze the late time behaviors of an unstable D-brane.
Section 4 is devoted to conclusion.

\section{An Exact Inhomogeneous Solution}

Our purpose in this work is to understand the late time behaviors
of the inhomogeneous tachyon condensation in terms of an exact solution.
We take a specific frame which gives
the vanishing momentum density over all space-time.

Equations of motion of the tachyon $T$ and gauge field $A_\mu$
in the action (\ref{dbia}) are written as
\bear
& & \partial_\mu\left(\gamma
\;C^{\mu\nu}_{\rm S}\;\partial_\nu T\right)
+{\cal T}_p\sqrt{-X}\;\frac{d V}{d T} = 0,
\label{te} \\
& &\partial_\mu\left( \gamma \;C^{\mu\nu}_{\rm A}
\right) = 0, \qquad (\mu,\, \nu = 0,1,2,\cdots p),
\label{ge}
\eear
where $C^{\mu\nu}_{\rm S}$ and $C^{\mu\nu}_{\rm A}$ are the symmetric and
anti-symmetric parts of the cofactor $C^{\mu\nu}$
of the matrix $(X)_{\mu\nu}$ in Eq.~(\ref{detX}), and we define
\bear\label{gamp}
\gamma \equiv \frac{{\cal T}_pV}{\sqrt{-X }}.
\eear

Conservation of energy-momentum is described by
\bear\label{cem}
\partial_\mu T^{\mu\nu} = \partial_\mu\left(\gamma\;
C^{\mu\nu}_{\rm S}  \right) =  0,
\eear
where $T^{\mu\nu}$ is the energy-momentum tensor.
Hamiltonian density ${\cal H}$ is expressed as
\bear \label{ham}
{\cal H} &=& \Pi_T\dot T + \Pi^i\dot A_i - {\cal L}
\nn \\
&=& \sqrt{\Pi^i\Pi^i + \Pi_T^2 + P_iP_i
+ (\Pi^i\partial_i T)^2
+ {\cal T}_p^2 V^2\det (h)} - A_0\partial_i \Pi^i,
\eear
where $\dot T \equiv \partial_0 T$, $h_{ij} = \delta_{ij}
+ \partial_i T\partial_j T + F_{ij}$, ($i,j = 1,2, \cdots p$),
the conjugate momenta, $\Pi_T$ and $\Pi^i$,
for the tachyon $T$ and gauge field $A_i$ respectively,
and the conserved linear momentum, $P_i$, associated
with the translation symmetry are
\bear
\Pi^i &\equiv& \frac{\delta S}{\delta \dot A_i} =
\gamma C_{{\rm A}}^{ti},
\nn \\
\Pi_T &\equiv&\frac{\delta S}{\delta \dot T}=
\gamma C_{{\rm S}}^{t\mu}\partial_\mu T,
\nn \\
P_i &=& \Pi^jF_{ji} -\Pi_T\partial_i T.
\label{cjm}
\eear

From now on, let us introduce an exponentially decreasing tachyon potential,
\bear \label{pot3}
V(T) =V_0\, e^{-T/R},
\eear
where $V_0$ and $R$ are arbitrary constants.
We consider an ansatz for fields which live on the worldvolume
of the unstable D$p$-brane,
\bear \label{ans1}
T= T(x^0,x^1), \quad E\equiv F_{02}(x^0, x^1),
\quad B\equiv F_{12}(x^0,x^1),
\eear
and, for simplicity, turn off all other components of the gauge fields.

Then the only non-vanishing linear momentum in Eq.~(\ref{cjm}) is
\bear\label{mmt}
P_1 &=& -\gamma\left(\dot T T'+ E B\right),
\eear
where $T'= \partial_1 T$.
Here, we choose the zero-momentum frame due to cancelation of
the effects of tachyon and electromagnetic fields,
\bear\label{vmmt}
EB = -\dot T T'.
\eear
Additionally the determinant $X$ in Eq.~(\ref{detX}) under
condition~(\ref{vmmt}) is factored as
\bear\label{detX2}
-X =(1+ T'^2 + B^2)(1-\dot T^2-E^2).
\eear
Conservation of energy-momentum, $\partial_\mu T^{\mu\nu}=0$, under the
condition (\ref{vmmt}) leads to an observation that
the energy density $T^{00}$ has only spatial dependence
and $T^{11}$ time dependence, i.e.,
\bear
T^{00} &=&  T^{00} (x) =
{\cal T}_pV \sqrt{\frac{1 + {T'}^2 + B^2}{1-\dot T^2-E^2}}\,,
\label{T00}\\
T^{11} &=&  T^{11} (t)=
- {\cal T}_pV \sqrt{\frac{1-\dot T^2-E^2}{1 + {T'}^2 + B^2}}\,,
\label{T11}
\eear
where we used the notations, $t=x^0$ and $x=x^1$.
The Eqs.~(\ref{T00}) and (\ref{T11}) are rewritten by
\bear
&&{\cal T}_p V(T) = f(t) g(x),
\label{fg1} \\
&& \frac{1-\dot T^2-E^2}{1 + {T'}^2 + B^2}=
\left(\frac{f(t)}{g(x)}\right)^2,
\label{fg2}
\eear
where
\bear\label{fg3}
f(t) \equiv  \sqrt{-T^{11}(t)},
\qquad g(x) \equiv  \sqrt{T^{00}(x)}.
\eear
Under the tachyon potential (\ref{pot3}),
the equation of motion for the tachyon field (\ref{te})
is simplified as
\bear
\frac{\ddot T}{1-\dot T^2 - E^2} - \frac{T''}{1  + T'^2 + B^2 }
- \frac{1}{R} = 0.
\label{te2}
\eear

Using the Eqs.~(\ref{fg1}), (\ref{fg2}), and (\ref{te2}),
we arrive at the following important results,
\bear\label{EB4}
E= E(t), \qquad B = B(x).
\eear
The derivation of this equation
is rather technical and therefore is recorded in Appendix.
The factorization (\ref{fg1}) implies
\bear
&&  \dot T(t) = - \frac{R \dot f(t)}{f(t)},
\qquad  T'(x) = - \frac{R g'(x)}{g(x)}.
\label{TdTp}
\eear
From the relations (\ref{vmmt}), (\ref{EB4}), and (\ref{TdTp})
we get
\bear
E(t) &=& \alpha \dot T = -\frac{\alpha R \dot f}{f},
\nn\\
B(x) &=& -\frac{1}{\alpha} T' = \frac{R g'}{\alpha  g},
\label{EB5}
\eear
where $\alpha$ is an arbitrary constant.
Inserting the expressions (\ref{TdTp}) and (\ref{EB5}) into
the Eq.~(\ref{fg2}), we obtain the first-order differential equations for
$f(t)$ and $g(x)$,
 \bear
\dot f^2 - \frac{f^2}{R^2(1 +\alpha^2)} +
\frac{ f^4}{\xi R^2(1 + \alpha^2)} = 0,
\label{feq} \\
g'^2 + \frac{\alpha^2  g^2}{R^2(1 + \alpha^2)}
- \frac{ \alpha^2  g^4}{\xi R^2(1 + \alpha^2)} = 0,
\label{geq}
\eear
where $\xi$ is a positive constant.
Solutions for the equations (\ref{feq}) -- (\ref{geq}) are given by
\bear
f(t) &=& \pm\sqrt{\xi}\,{\rm sech}
\left(\frac{ t-c_1}{R\sqrt{1+\alpha^2}}\right),
\label{fg4} \\
g(x) &=& \pm \sqrt{\xi}\, {\rm sec}\left(\frac{\alpha
(x-c_2)}{R \sqrt{1+\alpha^2}} \right),
\label{fg5}\eear
where $c_1$ and $c_2$ are integration constants, which represent
the translation symmetries along time and spatial directions respectively.
Of course, the expressions (\ref{feq}) -- (\ref{fg5}) satisfy
the $2$-component of the gauge equation (\ref{ge}),
\bear\label{ge2}
g^2 \dot E =  f^2 B'.
\eear
Substituting the solution (\ref{fg4}) into the Eq.~(\ref{gamp}),
we find
\bear\nn
\gamma = \frac{{\cal T}_p V}{\sqrt{-X}} =
\frac{{\cal T}_p V}{\sqrt{(1 + T'^2 + B^2)(1 - \dot T^2 - E^2)}}
= \xi.
\eear
Finally we obtain an exact solution for the tachyon field by inserting
the expression (\ref{fg4}) into (\ref{fg1}),
\bear\label{tcn2}
T(t,x) = R \ln\left[ \frac{{\cal T}_p V_0}{\gamma}\,
{\rm cosh} \left(\frac{t-c_1}{R \sqrt{1+\alpha^2}} \right)
\left|{\rm cos}
\left(\frac{\alpha  (x-c_2)}{R \sqrt{1+\alpha^2}}
\right)\right| \right].
\eear
This solution is characterized by two
parameters, $\gamma$ and $\alpha$.
We will investigate the roles of these parameters in section 3.
At first glance this result seems to be unnatural since $T(t,x)$
has the periodic divergencies in the limit
$\left|\cos\left(\frac{\alpha (x-c_2)}{R\sqrt{1+\alpha^2}}\right)
\right| \to 0$
due to the property of cosine function.
Actually in the spatially periodic regions at the initial time $t=0$
which satisfy
\bear\label{inicon}
\left|\cos\left(\frac{\alpha (x-c_2)}{R\sqrt{1+\alpha^2}}
\right)\right|
<  \frac{(\gamma / {\cal T}_p V_0)}{
\cosh\left(\frac{c_1}{R\sqrt{1+\alpha^2}}\right)} <1,
\eear
the corresponding tachyon field $T(t=0, x)$ is negative.
In these regions the DBI-type effective action does not provide
a good description for the dynamics of an unstable D-brane as we
explained in the section 1.
In order to describe the late time behaviors of the decay process
of an unstable D-brane, we restrict our interest to the spatially
periodic regions which correspond to the large positive value
of tachyon field. We will describe the details in the next section.

\section{Late Time Behaviors of the Decay Process}

It was observed in the previous section that there is an exact
solution (\ref{tcn2}) for the exponentially decreasing tachyon potential
in momentum zero frame.
Our purpose in this section is  to analyze the
solution (\ref{tcn2}) in superstring theory.
Since a total charge is conserved
(we will mention later in detail at subsection 3.3)
and the tachyon potential has ${\rm Z}_2$-symmetry under $T \to -T$,
and runaway property $V(\pm\infty)=0$,
we employ a tachyon potential composed of two parts,
\bear \label{tcpot2}
V(T)= V_0\, e^{-\frac{|T|}{\sqrt{2}}}=\left\{\begin{array}{lclcc}
V_0\, e^{-\frac{T}{\sqrt{2}}}& ~& T > 0 & ~ &({\rm I})\\
V_0\, e^{\frac{T}{\sqrt{2}}}& ~ & T < 0 & ~ & ({\rm II})
             \end{array}\right..
\eear

Tachyon profile is read from the Eq.~(\ref{tcn2})
in the regions (I) and (II) by choosing the appropriate
integration constants, $c_1$ and $c_2$,
\bear \label{tcn3}
T(t,x) = \left\{\ba{lcc}
\sqrt{2}\ln \left[ \frac{{\cal T}_pV_0}{\gamma}\,
{\rm  cosh} \left( \frac{ t+t_0}{\sqrt{2}\sqrt{1+\alpha^2}} \right)
\cos\left( \frac{\alpha x_{\rm I} } {\sqrt{2}\sqrt{1+\alpha^2}}
-\frac{\pi}{2}\right)\right], &~&({\rm I}) \\
-\sqrt{2}\ln \left[ \frac{{\cal T}_pV_0}{\gamma}\,
{\rm  cosh} \left( \frac{ t+t_0}{\sqrt{2}\sqrt{1+\alpha^2}} \right)
\cos\left( \frac{\alpha x_{{\rm II}} }{\sqrt{2}\sqrt{1+\alpha^2}}
+\frac{\pi}{2}\right)\right], &~&({\rm II})\ea\right.,
\eear
where $x_{\rm I}$($x_{{\rm II}}$) represents the spatial coordinate
belonging to the region (I)((II)),
$t_0$ is some large value introduced to figure out the late
time behaviors of the inhomogeneous fields.
The ranges of $x_{\rm I}$ and $x_{{\rm II}}$ are given by
\bear\label{vlr}
&& x^{-}_{4m+1} \le x_{\rm I} \le x^{+}_{4m+1},
\nn \\
&&  \tilde{x}^{-}_{4m+1} \le x_{{\rm II}} \le \tilde{x}^{+}_{4m+1},
\quad (m=\mbox{integer}).
\eear
where $x^{\pm}_{4m+1}$ and $\tilde{x}^{\pm}_{4m+1}$ denote
\bear\label{xtr}
x^{\pm}_{4m+1}&=&\frac{\sqrt{2}\sqrt{1 + \alpha^2}}{\alpha}
\left[\pm\cos^{-1}\left(\frac{\gamma}{{\cal T}_pV_0}\right)
+\frac{(4m+1)\pi}{2}\right], \nn \\
\tilde{x}^{\pm}_{4m+1}&=&\frac{\sqrt{2}\sqrt{1 + \alpha^2}}{\alpha}
\left[\pm\cos^{-1}\left(\frac{\gamma}{{\cal T}_pV_0}\right)
-\frac{(4m+1)\pi}{2}\right],
\quad
( 0\le \gamma\le {\cal T}_pV_0).
\eear
There are periodic regions, where the decay process
of the unstable D-brane is not described well by the
solution (\ref{tcn3}), referred as inaccessible regions,
\bear
\tilde x^{+}_{4m + 1} < x_{{\rm inaccessible}} <  x^{-}_{4m + 1},
\quad \mbox{and}\quad
 x^{-}_{4m + 1} < x_{{\rm inaccessible}} < \tilde x^{+}_{4m + 5}.
\eear

To illustrate the solution (\ref{tcn3}) graphically, we draw
two figures for the tachyon potential $V(T)$ and tachyon field
$T(t,x)$ in Fig.1.
The arrows in Fig.1 (b) represent growing tachyon field as time elapses.
The corresponding tachyon fields spans the ranges at initial time $t=0$,
\bear
T_0 \le T_{{\rm I}}(t,x) \le T_{{\rm max}},
\qquad
-T_{{\rm max}} \le T_{{\rm II}}(t,x) \le -T_0 ,
\eear
where
\bear
&&T_0 = \sqrt{2}\ln \cosh\left( \frac{t_0}{\sqrt{2}\sqrt{1
+ \alpha^2}}\right),
\nn \\
&&T_{{\rm max}} = \sqrt{2}\ln \left[ \frac{{\cal T}_pV_0}{\gamma}\,
{\rm  cosh} \left( \frac{t_0}{\sqrt{2}\sqrt{1+\alpha^2}}\right)\right].
\nn\eear
\begin{figure}[t]
\centerline{\epsfig{figure=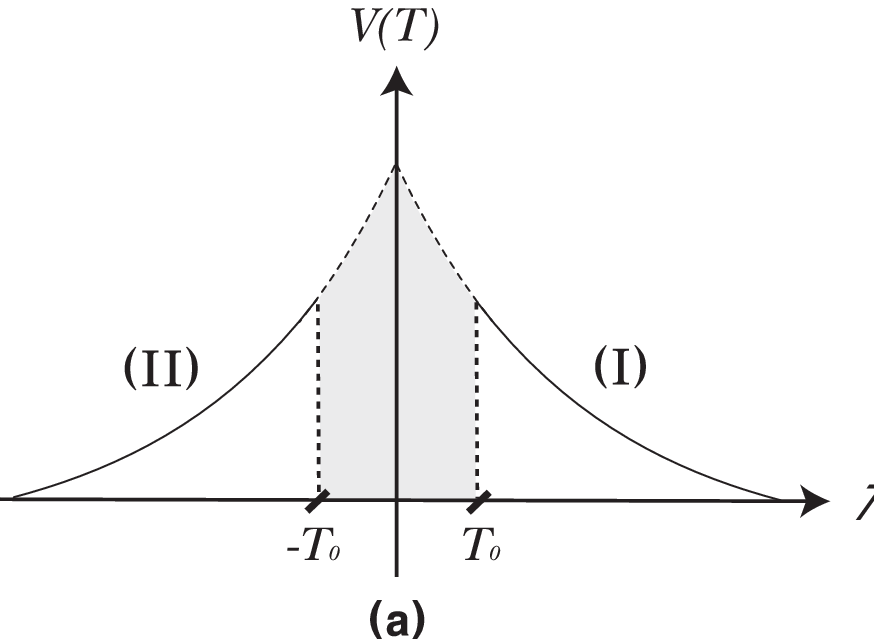,height=60mm}
\hfill
\epsfig{figure=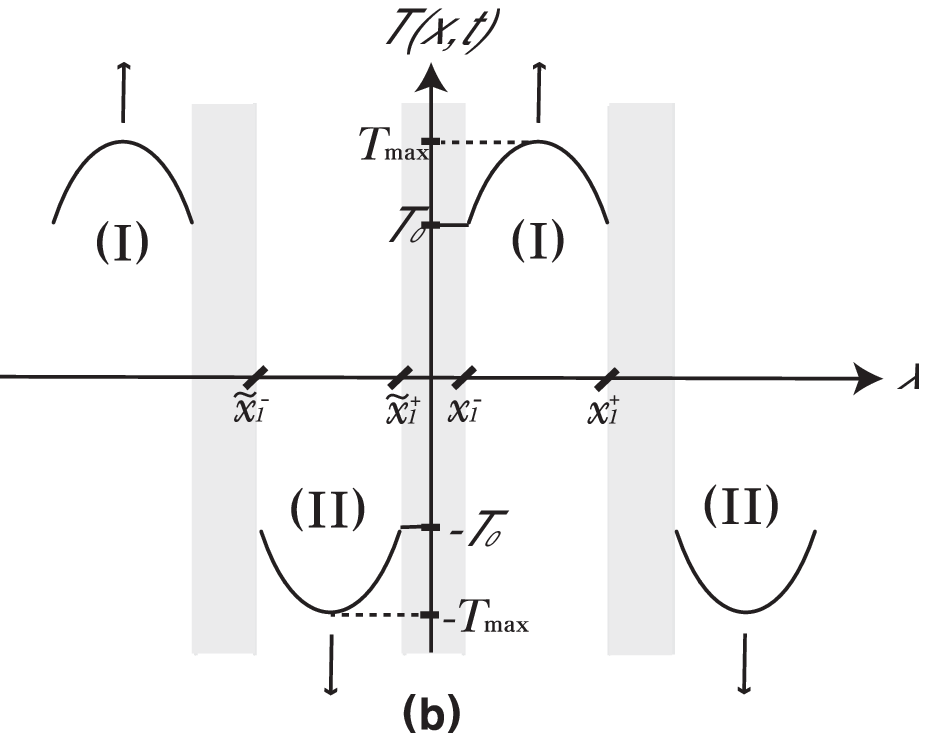,height=73mm}}
\caption{(a) Shape of $V(T)=e^{\frac{-|T|}{\sqrt{2}}}$ for $|T| \ge T_0$.
(b) Profiles of tachyon field $T(x,t)$. Shaded regions
represent the inaccessible regions.}
\end{figure}

Electromagnetic fields in Eq.~(\ref{EB5})
take the functional forms
\bear
E(t)&=& \alpha \dot T(t) = \left\{\ba{lcc}
\frac{\alpha}{\sqrt{1 + \alpha^2}}\tanh \left( \frac{ t+t_0}{
\sqrt{2}\sqrt{1 + \alpha^2}}\right),
&~& ({\rm I}) \\
-\frac{\alpha}{\sqrt{1 + \alpha^2}}\tanh \left( \frac{ t+t_0}{
\sqrt{2}\sqrt{1 + \alpha^2}}\right),
&~& ({\rm II}) \ea\right. ,
\label{EE1} \\
B(x) &=& -\frac{1}{\alpha}T'(x)=
\left\{\ba{lcc}
\frac{1}{\sqrt{1 + \alpha^2}}\tan\left(\frac{\alpha x_{\rm I}}{
\sqrt{2}\sqrt{1 + \alpha^2}}-\frac{\pi}{2}\right),
&~&({\rm I}) \\
-\frac{1}{\sqrt{1 + \alpha^2}}\tan\left(\frac{\alpha x_{{\rm II}}}{
\sqrt{2}\sqrt{1 + \alpha^2}}+\frac{\pi}{2}\right),
&~&({\rm II})\ea\right..
\label{BB1}
\eear
The configuration for the tachyon field (\ref{tcn3}) represents the time
evolution of the spatially periodic profiles governed by two
parameters, $\gamma$ and $\alpha$.
$\gamma$ adjusts the distribution of the energy density, while $\alpha$
is the scaling parameter of time and controls the period.
We investigate the roles of the parameter $\alpha$.

\subsection{$\alpha \to 0$ case : Homogeneous rolling tachyon}

When $\alpha$ goes to zero, the period of tachyon profile
approaches infinity.
Thus the corresponding solution in Eq.~(\ref{tcn3})
describes the homogeneous rolling tachyon, i.e., the spatial
inhomogeneity is neglected.
In the region (I) for $m=0$ in Eq.~(\ref{vlr}),
the Eqs.~(\ref{tcn3}), (\ref{EE1}), and (\ref{BB1})
are given by
\bear
T(t,x) &=& \sqrt{2} \ln\left[\frac{{\cal T}_pV_0}{\gamma}
\cosh\left(\frac{t+t_0}{\sqrt{2}}\right)
\cos\left(\frac{\alpha x_{\rm I}}{\sqrt{2}}\right)\right],\nn \\
B(x)&=&\tan\left(\frac{\alpha x_{\rm I}}{\sqrt{2}}\right),\nn \\
\dot T(t) &=& \tanh\left(\frac{t+t_0}{\sqrt{2}}\right),
\nn \\
T'(x) &=& E(t) = 0
\label{hrot}
\eear
with
\bear \nn
-\sqrt{2} \cos^{-1}\left(\frac{{\cal T}_pV_0}{\gamma}\right)
\leq \alpha x_{\rm I}\leq \sqrt{2} \cos^{-1}\left(
\frac{{\cal T}_pV_0}{\gamma}\right).
\eear
This analysis is also applicable to the case of the region (II).
It is well-known that when the tachyon and
electromagnetic fields in the DBI-type effective action (\ref{dbia})
depend on only one spacetime coordinate,
the results of equations of motion for the given system show that
all the electromagnetic fields are constants~\cite{Kim:2003in,Kim:2003he}.
In this limit, the period of magnetic field $B(x)$ reaches to infinity
but its amplitude remains finite. Therefore
the results in Eqs.~(\ref{hrot}) represent a homogeneous rolling tachyon
with almost constant magnetic field and constant energy
density $T^{00} = {\cal T}_p^2V_0^2/\gamma$.

The pressure and tachyon matter density are
\bear
&&T^{11}(t) = T^{22} = -\gamma\, {\rm sech}^2\left(\frac{ t+t_0}{
\sqrt{2}\sqrt{1 + \alpha^2}}\right),
\nn \\
&&\Pi_T(t,x) = \frac{{\cal T}_p^2}{\gamma}
 \tanh\left(\frac{t+t_0}{\sqrt{2}}\right).
\eear
In the limit of $t\to \infty$, we obtain the pressureless matter with constant
energy density and tachyon matter density~\cite{Sen:2002an}.

\subsection{$\alpha \to \infty$ case : Static configurations}\label{tachkink}

On the other hand, when $\alpha$ goes to infinity, $x^{\pm}_{4m+1}$
and $\tilde{x}^{\pm}_{4m+1}$ reach to fixed finite values,
\bear\label{xtrr}
x^{\pm}_{4m+1}&=&\sqrt{2}
\left[\pm\cos^{-1}\left(\frac{\gamma}{{\cal T}_pV_0}\right)
+\frac{(4m+1)\pi}{2}\right], \nn \\
\tilde{x}^{\pm}_{4m+1}&=&\sqrt{2}
\left[\pm\cos^{-1}\left(\frac{\gamma}{{\cal T}_pV_0}\right)
-\frac{(4m+1)\pi}{2}\right].
\eear
From the Eq.~(\ref{tcn3}) we can easily check that
the time dependence disappears and the configuration of tachyon
becomes a static solution in this limit.
The expressions of the Eqs.~(\ref{tcn3}), (\ref{EE1}), and (\ref{BB1})
at finite time $t << \alpha $ in the region (I) are given by
\bear
T(t,x) &=& \ln\left[\frac{{\cal T}_pV_0}{\gamma} \cos\left(
\frac{x_{\rm I}}{\sqrt{2}}-\frac{\pi}{2}\right)\right],
\nn \\
T'(x) &=& -\tan \left(\frac{x_{\rm I}}{\sqrt{2}}-\frac{\pi}{2}\right),
\nn \\
E(t) &=& \dot T(t) = B(x) = 0,
\label{tk}
\eear
where $\gamma$ adjusts the distribution of energy density for
the given static configuration.
In the limit $\gamma\to 0$, most of energy is localized at
$x^{\pm}_{4m+1}$.
The energy stored in half period of one cycle in the limit
is obtained by
\bear\label{decn}
\left(\int_{-x_0}^{\tilde x^+_1} dx
+ \int_{x^-_1}^{x_0} dx \right)\, T^{00}(x)|_{\gamma\to 0}
= 2\sqrt{2}\, {\cal T}_pV_0,
\eear
where $x_0= \sqrt{1 + \alpha^2}\pi/\sqrt{2}\alpha$.
Since our exact solution is valid only for large $T$,
the decent relation (\ref{decn}) is approximately correct.

As we have worked in the above subsections 3.1 and 3.2, $\alpha$ is the only
parameter which governs the time scale and period along spatial
direction for a given $\gamma$ in the solution (\ref{tcn3}).
Thus $\alpha$ is an interpolating parameter between
the homogeneous rolling tachyon ($\alpha=0$) and the static
configuration ($\alpha \to \infty$).

\subsection{Late time behaviors}

As time elapses, the tachyon profiles (\ref{tcn3}) in the regions
(I) and (II) grow up along the arrows in Fig.1 (b).
In the  interesting regions (\ref{vlr}),
all physical quantities can be expressed
explicitly, and the unstable system evolves without singularity
in our system.
Non-vanishing components of energy-momentum tensor are given by
\bear T^{00}(x)&=& \left\{\ba{lcc} \gamma\sec^2\left(\frac{\alpha
x_{\rm I}}{ \sqrt{2}\sqrt{1 + \alpha^2}}-\frac{\pi}{2}\right),
&~&({\rm I}) \\
\gamma\sec^2\left(\frac{\alpha x_{{\rm II}}}
{\sqrt{2}\sqrt{1 + \alpha^2}}+\frac{\pi}{2}\right),
&~&({\rm II})\ea\right.,
\label{Ttt3} \\
T^{11}(t) &=& -\gamma\, {\rm sech}^2\left(\frac{ t+t_0}{
\sqrt{2}\sqrt{1 + \alpha^2}}\right) \,\, {\rm for}\,\, ({\rm I})\,\,
{\rm and} \,\, ({\rm II}),
\label{Txx3} \\
T^{22}(t,x) &=& \left\{\ba{lcc}
-\gamma\left[1-\frac{1}{1 + \alpha^2}\tanh^2
\left(\frac{t+t_0}{\sqrt{2}\sqrt{1 + \alpha^2}}\right) +
\frac{\alpha^2}{1 + \alpha^2}\tan^2
\left(\frac{\alpha x_{\rm I}}{\sqrt{2}\sqrt{1 + \alpha^2}}-\frac{\pi}{2}\right)\right],
&~&({\rm I}) \\
-\gamma\left[1-\frac{1}{1 + \alpha^2}\tanh^2
\left(\frac{t+t0}{\sqrt{2}\sqrt{1 + \alpha^2}}\right) +
\frac{\alpha^2}{1 + \alpha^2}\tan^2
\left(\frac{\alpha x_{{\rm II}}}{\sqrt{2}\sqrt{1 + \alpha^2}}
+\frac{\pi}{2}\right)\right],
&~&({\rm II}) \ea\right..
\nn \\ \label{Tyy3}
\eear
Since the momentum density flow is zero in the interesting regions of
the solution (\ref{vlr}),
the initial energy density distribution is not changed
in time direction.
As time goes to infinity, pressure along the inhomogeneous direction,
$x$, goes to zero exponentially~\cite{Sen:2002an}.
$T^{22}$ component depends on $t$ and $x$-coordinates
for finite $\alpha$. However, in $\alpha\to 0$ limit
(homogeneous rolling tachyon limit),
$T^{22}$ and $T^{11}$ share with the same behavior
in time direction.

The solution (\ref{tcn3}) also provides the expression for the
electric flux density which satisfies Gauss constraint
$\partial_i \Pi^i =0$ and the gauge equation (\ref{ge2}),
\bear
\Pi_2 &=& \alpha \Pi_T =
\left\{\ba{lc}
\frac{\alpha\gamma}{\sqrt{1 + \alpha^2}}\tanh \left( \frac{ t+t_0}{
\sqrt{2}\sqrt{1 + \alpha^2}}\right)\sec^2
\left(\frac{\alpha x_{\rm I}}{\sqrt{2}\sqrt{1 + \alpha^2}}
-\frac{\pi}{2}\right),
& ({\rm I}) \\
-\frac{\alpha\gamma}{\sqrt{1 + \alpha^2}}\tanh \left( \frac{ t+t_0}{
\sqrt{2}\sqrt{1 + \alpha^2}}\right)\sec^2
\left(\frac{\alpha x_{{\rm II}}}{\sqrt{2}\sqrt{1 + \alpha^2}}
+\frac{\pi}{2}\right),
& ({\rm II}) \ea\right..
\label{fsd}
\eear
This expression denotes that absolute values of
the electric flux and tachyon matter densities increase
in the D-brane decay process.
However, total string charge accumulated in the interesting regions
is conserved in one period~;
\bear
\left(\int_{\tilde{x}^{-}_{4m+1}}^{\tilde x^{+}_{4m+1}} dx
+\int_{x^{-}_{4m+1}}^{ x^{+}_{4m+1}} dx\right) \, \Pi_2(t,x) = 0.
\eear

Combining the Eqs.~(\ref{EE1}), (\ref{Ttt3}), and (\ref{fsd}),
we obtain the following two relations,
\bear
&& 1 - \dot T^2 - E^2 = {\rm sech}^2\left(\frac{ t+t_0}{
\sqrt{2}\sqrt{1 + \alpha^2}}\right),
\label{ETd}\\
&& \Pi_2^2 + \Pi_T^2 = {\cal H}^2
\tanh^2 \left( \frac{ t+t_0}{ \sqrt{2}\sqrt{1 + \alpha^2}}\right),
\label{PPH}
\eear
where the Hamiltonian density ${\cal H}$ is given by
\bear\label{Ham2}
{\cal H}= \sqrt{\Pi_2^2 + \Pi_T^2 + {\cal T}_p^2 V^2
(1 + T'^2 + B^2)} = T^{00}(x).
\eear

Since there is no singularity in interesting regions
in time direction, the obtained solution describes safely
the decay processes near the tachyon vacuum ($V\to 0$)
in $t\to \infty$ limit.
As time goes to infinity, the time dependent electric fields in
the regions (I) and (II) become constants with opposite sign,
\bear
E(t)|_{T\to\infty}=
\alpha\dot T(t)|_{T\to\infty}
=  \left\{\ba{lcc} \frac{\alpha}{\sqrt{1 + \alpha^2}},
&~& ({\rm I}) \\
-\frac{\alpha}{\sqrt{1 + \alpha^2}},
&~& ({\rm II}) \ea\right..
\label{EE2}
\eear
These relations in tachyon vacuum ($V\to 0$) reproduce the well-known
expression~\cite{Mukhopadhyay:2002en,Gibbons:2002tv},
\bear\label{TE1}
&& 1 - \dot T^2 - E^2 \to 0.
\eear
This result leads to an intriguing observation.
As we have discussed in subsection \ref{tachkink},
the system gives a static configuration in the limit $\alpha\to\infty$.
At initial time the electric field is equal to zero in Eq.~(\ref{tk}).
As time evolves to infinity ($t/\alpha\to\infty$),
the electric field becomes critical and $\dot T$ is suppressed to zero,
\bear
|E(t)| \to 1, \qquad |\dot T(t)| \to 0.
\eear

As we have seen in Eq.~(\ref{Ham2}), the energy density is composed of
three parts, such as string flux density ($\Pi_2$),
tachyon matter density ($\Pi_T$), and tachyon potential energy.
As time evolves, the contributions from $\Pi_2$ and
$\Pi_T$ increase,
while the contribution from the tachyon potential decreases.
Finally the unstable D-brane disappears at the tachyon vacuum ($V\to 0$).
The resultant energy density in the tachyon vacuum is composed of two
parts~\cite{Gibbons:2000hf,Gibbons:2002tv},
\bear \label{pph}
{\cal H}^2=\Pi_2^2 + \Pi_T^2,
\eear
where
\bear
\Pi_2(t,x)|_{T\to\infty} =
\alpha \Pi_T(t,x)|_{T\to\infty}
= \left\{\ba{lc}
\frac{\alpha\gamma}{\sqrt{1 + \alpha^2}}\sec^2
\left(\frac{\alpha x_{\rm I}}{\sqrt{2}\sqrt{1 + \alpha^2}}-\frac{\pi}{2}\right),
& ({\rm I}) \\
-\frac{\alpha\gamma}{\sqrt{1 + \alpha^2}}\sec^2
\left(\frac{\alpha x_{{\rm II}}}{\sqrt{2}\sqrt{1 + \alpha^2}}
+\frac{\pi}{2}\right),
& ({\rm II}) \ea\right..
\label{PP}
\eear

\section{Conclusion}

We have investigated the spatially inhomogeneous decay of an unstable D-brane
in DBI-type effective action.
We found an exact solution under an exponentially
decreasing tachyon potential.
The resulting solution involves the periodic inaccessible region
along the inhomogeneous direction,
while the behavior in time direction is well-defined.
The solution is governed by two parameters, $\gamma$ and $\alpha$.
$\gamma$ adjusts the distribution of energy density,
and $\alpha$ is an interpolating parameter between the homogeneous
rolling tachyon and the static solution.

It is well-known that the inhomogeneous rolling tachyon with a runaway
type tachyon potential forms caustics with multi-valued regions beyond
a finite critical time.
After the critical time the unstable
system may not be described by DBI-type tachyon effective action.
However, as we have seen in section 3,
it was possible to describe the late time behaviors
of an unstable D-brane in the interesting regions due to
the nontrivial roles of the spacetime dependent
electromagnetic fields.
Therefore our solution may open a possibility
to find the caustic free tachyon field solution in a specific setting
with spacetime dependent electromagnetic fields
in tachyon effective field theory.

As time evolves, all physical quantities are well defined and go to
tachyon vacuum ($V=0$) without developing further singularities in
the interesting regions. Electric flux density, which is proportional to
the tachyon matter density with a constant ratio $\alpha$, increase
in magnitude, but have the opposite signs in the region (I) and
(II). They finally reach to space dependent finite configurations.
As a result, the energy stored in the unstable D-brane at the
initial stage is converted to that of the string fluid and the
tachyon matter. Since two interesting regions in one cycle (See Fig.1 (b))
go to the different vacua ($T\to\infty$ and $T\to-\infty$) in
$t\to\infty$ limit, inaccessible region
 between them contains a
topological kink which seems to be interpreted as D$(p-1)\bar {\rm
D}(p-1)$ or D$(p-1){\rm F}1 \bar {\rm D}(p-1){\rm F}1$ .

\section*{Acknowledgements}

We would like to thank Piljin Yi for valuable discussions
and Yoonbai Kim for many helpful comments on the manuscript.
C.L. is grateful to the department of physics at Sungkyunkwan university
for hospitality during staying there for the completion of this work.
This work was supported by Korea Science $\&$ Engineering Foundation
(KOSEF R01-2004-000-10526-0 for C.L.),
and is the result of research activities (Astrophysical Research
Center for the Structure and Evolution of the Cosmos (ARCSEC))
supported by Korea Science $\&$ Engineering
Foundation~(O.K.).

\section*{Appendix: Derivation of $E=E(t)$ and $B=B(x)$}
\makeatletter
\renewcommand{\theequation}{A.\arabic{equation}}
\@addtoreset{equation}{section}
\makeatother
\setcounter{equation}{0}
In this appendix we present detailed derivation of $E(t)$ and $B(x)$.
From the Eqs.~(\ref{fg1}), (\ref{fg2}), and (\ref{te2}), we obtain
\bear
E^2 &=& 1- \dot T^2 - R\ddot T
+ R f^2\, \rho (x),
\nn \\
B^2 &=& - \left(1 + T'^2 + RT''\right)
 + R g^2\,\sigma (t),
\label{BB}\eear
where
\bear
\rho (x) \equiv \frac{T''}{g^2},
\qquad
\sigma (t) \equiv \frac{\ddot T}{ f^2}.
\eear
Differentiations of the two equations in Eq.~(\ref{BB})
with respect to $x$ and $t$ yield
\bear
2 EE' &=& R f^2\, \rho' (x),
\nn \\
2 B \dot B &=&R g^2\,\dot \sigma (t).
\label{EBd}\eear
Product of the two equations in Eq.~(\ref{EBd}) gives
\bear\label{EBd2}
E'^2 =\dot B^2=  - \frac{f^3 g^3}{4 \dot f g'} \rho'\dot\sigma,
\eear
where we used the Eqs.~(\ref{vmmt}), (\ref{TdTp}), and
Bianchi identity for the gauge field strength,
\bear\label{Bian}
E' = \dot B.
\eear
The relations in Eqs.~(\ref{EBd}) and (\ref{EBd2}) imply that $E$, $B$, $E'$,
and $\dot B$ are separable functions of $t$ and $x$.
The separable properties for the electromagnetic fields, $E$ and $B$,
in Eq.~(\ref{BB}) impose on the conditions for the functions,
$f(t)$ and $g(x)$,
\bear
&&1- \dot T^2 - R \ddot T = c  f^2,\,
\nn\\
&& 1 + T'^2 + R T''=  d  g^2,
\label{sepc}
\eear
where $c$ and $d$ are arbitrary constants.
Insertion of the relations (\ref{sepc}) into the Eq.~(\ref{BB})
produces
\bear\label{sep2}
E^2B^2= f^2g^2(c + R \rho (x))(-d + R\sigma (t))
 = \frac{R^4 \dot f^2 g'^2}{f^2g^2}= \dot T^2 T'^2.
\eear
The Eq.~(\ref{sep2}) is separated into two parts,
\bear\label{sep3}
R\ddot T &=&\frac{\dot T^2}{\zeta}+ d f^2,
\nn \\
RT''&=& \zeta T'^2 - c g^2,
\eear
where $\zeta$ is an arbitrary positive constant.
The relations between Eqs.~(\ref{sepc}) and (\ref{sep3}) give
\bear \label{sep4}
&&\left(1 + \frac{1}{\zeta}\right) \dot T^2 + ( c+ d) f^2 -1 = 0,
\nn \\
&& ( 1 + \zeta ) T'^2 - ( c+ d) g^2 + 1 = 0.
\eear
Differentiating the relations in Eq.~(\ref{sep4}), we find
\bear \label{sep5}
\sigma (t) &=& \frac{\ddot T}{f^2} =
\frac{\zeta (c + d)}{R(1 + \zeta)} = \mbox{constant},
\nn \\
\rho (x) &=& \frac{T''}{g^2} =
-\frac{c + d}{R(1 + \zeta)} = \mbox{constant}.
\eear
Using the Eqs.~(\ref{BB}) and (\ref{sep5}), we finally obtain
\bear
E= E(t), \qquad B = B(x).
\eear

\end{document}